\documentclass[aps,prl,reprint,letterpaper,superscriptaddress,showpacs]{revtex4-1}
\usepackage{amssymb}
\usepackage{keyval}%
\usepackage{graphicx}
\usepackage{dcolumn}
\usepackage{bm}
\usepackage{color}
\usepackage{amssymb}
\usepackage{amsmath}

\newcommand{\red}[1]{#1}

\newcommand{\ignore}[1]{}
\newcommand{\notextit}{}
\newcommand{\bfa}{Ba(Fe$_{1-x}$Co$_x$)$_2$As$_2$}

\ignore{
\pagestyle{fancy} %
\pagestyle{fancyplain} %
\lhead{PRL \textbf{112}, to appear (2014)} %
\chead{PHYSICAL REVIEW LETTERS\vspace{-2pt}}
\rhead{\small MS No. \\ LF14168} %
\cfoot{\sc\thepage} %
\lfoot{} %
\rfoot{}
}

\begin{document}

\title{
%Strong coupling of the charge redistribution on the iron $3d$ orbitals and the anion polarization in \bfa}
\red{Strong Coupling of the Iron-Quadrupole and Anion-Dipole Polarizations in \bfa}}
\author{Chao Ma}
%\email{cma@iphy.ac.cn}
\affiliation{Condensed Matter Physics and Materials Sciences Department, Brookhaven National Laboratory, Upton, New York 11973, USA\\}
\affiliation{Beijing National Laboratory for Condensed Matter Physics, Institute of Physics, Chinese Academy of Sciences, Beijing 100190, China\\}
\author{Lijun Wu}
\author{Wei-Guo Yin}
\email{wyin@bnl.gov}
\affiliation{Condensed Matter Physics and Materials Sciences Department, Brookhaven National Laboratory, Upton, New York 11973, USA\\}
\author{Huaixin Yang}
\author{Honglong Shi}
\author{Zhiwei Wang}
\author{Jianqi Li}
\affiliation{Beijing National Laboratory for Condensed Matter Physics, Institute of Physics, Chinese Academy of Sciences, Beijing 100190, China\\}
\author{C. C. Homes}
\author{Yimei Zhu}
\email{zhu@bnl.gov}
\affiliation{Condensed Matter Physics and Materials Sciences Department, Brookhaven National Laboratory, Upton, New York 11973, USA\\}

\date{Received 10 June 2013; accepted by PRL}

\begin{abstract}
We use a quantitative convergent beam electron diffraction (CBED) based method to image the valence electron density distribution in \bfa. We show a remarkable increase in both the charge quadrupole of the Fe cations and the charge dipole of the arsenic anions upon Co doping from $x=0$ ($T_c=0$~K) to $x=0.1$ ($T_c=22.5$~K). Our data suggest that an unexpected electronic correlation effect, namely strong coupling of Fe orbital fluctuation and anion electronic polarization, is present in iron-based superconductors.
\end{abstract}


\maketitle
%\thispagestyle{fancy}
%\narrowtext

\sloppy

Iron-pnictide high-temperature superconductivity develops when the `parent' spin-ordered \cite{Cruz} and likely orbital-ordered \cite{Kruger,Yin:PRL99,Lv,Chen} phase is suppressed, typically by introduction of dopant atoms \cite{review:Paglione}.
This phenomenon is similar to cuprate superconductivity but the orbital physics is new. Whether strong orbital fluctuation \red{(electronic oscillation among the Fe $3d$ orbitals)}
develops with the doping and creates a novel mechanism for high-$T_c$ superconductivity has been of great interest \cite{Chuang,Chu,Yi,Fisher,Kasahara,Nakajima,Stanev,Fernandes,Goto,Yoshizawa,Arham,Kim,Kontani,Yanagi}.
Strong orbital fluctuation arguably can be induced by phonons (iron \cite{Kontani} or arsenic \cite{Yanagi} atomic oscillation) and mediate the $s_{++}$-wave state, in contrast with the ``conventional'' spin fluctuation-mediated $s_\pm$-wave state \cite{review:Mazin,Christianson}. This problem remains outstanding because direct experimental evidence is lacking. Whereas, orbital fluctuation in principle can be directly measured as a dynamical change in the Fe charge quadrupole \cite{Kontani,Goto} and its signal can be enhanced by impurities \cite{Inoue}.

Concerning the charge multipole dynamics, there is another unique feature of iron-based superconductors, namely the unusually high electronic polarizability of the anions bonded to the Fe cations \cite{Sawatzky,Berciu}. Take \bfa\ \cite{Sefat} as a prototypical example \cite{review:Paglione,Chuang,Chu,Yi,Fisher,Kasahara,Nakajima,Stanev,Fernandes,Goto,Yoshizawa,Arham,Kim}. The electronic polarizability of As$^{3-}$ is $9-12$~{\AA}$^3$, which is much larger than $0.5-3.2$~{\AA}$^3$ of O$^{2-}$ \cite{Sawatzky}. In addition, the anions sit considerably out of the Fe plane (the Fe-As-Fe angle is $\sim71^\circ$). Thus the anions' electron clouds can be readily shifted by the electric field of the Fe charge monopole. Two effects of this Fe monopole-anion dipole interaction have been considered: (i) It weakens the effective electron-electron repulsion on Fe $3d$ orbitals~\cite{Sawatzky}, driving the system away from the Mott insulator regime \cite{Yang}. This effect has been taken into account in most theories for iron-based superconductivity, in which the anion degree of freedom is then assumed to be inactive. (ii) However, it was illustrated in a one-orbital model that the electronic oscillation of the anions could dress the electrons moving in the Fe plane to form `electronic polarons' and mediate them to form Cooper pairs \cite{Sawatzky,Berciu}. The arsenic orbitals were recently reported to be relevant to superconductivity in a spectroscopic ellipsometry study \cite{Charnukha}. Yet, there is also a lack of direct experimental evidence of those active roles of anion polarization. In any case, the coupling of the anion-dipole polarization and the Fe-orbital fluctuation \red{(or even the static Fe quadrupole polarization)} has not been discussed.

In this Letter, we use a quantitative convergent beam electron diffraction (CBED) based method \cite{Zuo,Wu99,Wu,Zhu} to image the valence electron density distribution in \bfa\ single crystals and thus obtain the information about the Co concentration dependence of the Fe charge quadrupole and the anion charge dipole. This method has been successfully applied to describe the valence electron distribution and essential bonding in various systems ranging from copper oxides \cite{Zuo,Wu99} to MgB$_2$ \cite{Wu} and to CaCu$_3$Ti$_4$O$_{12}$ \cite{Zhu}, etc. Here, we show a remarkable increase in the Fe charge quadrupole in \bfa\ upon Co doping from $x=0$ ($T_c=0$~K) to $x=0.1$ ($T_c=22.5$~K). Unexpectedly, we also observe an obvious boost in the charge dipole of the arsenic anions. Our data reveal a strong coupling between the \red{static} Fe quadrupole and anion dipole, and suggest the existence of a novel electronic correlation effect, namely strong Fe orbital fluctuation can be induced by the optical oscillation of the anions' electron cloud and vice versa, in iron-based superconductors.

The single crystals used in the present study were prepared by the high-temperature solution method using FeAs as flux \cite{Ma}.
For high-energy electron probes we can easily find regions with perfect crystal integrity that satisfy the requirement of charge density measurement. The specimens used for our transmission electron microscopy (TEM) observations were prepared by slicing off a thin sheet, a few tens of microns thick, from the single crystal, and followed by low-energy ion milling. The CBED experiments were performed at room temperature, using the JEOL-2200MCO TEM equipped with two aberration correctors and an in-column omega filter. The electron probe was focused within $\sim20$~nm sized area at a constant thickness (60-100 nm for different areas).
In addition, we carried out electron energy loss spectroscopy (EELS) on the FEI Tecnai F20 TEM equipped with a Gatan imaging filter to verify the information about Fe charge monopole. The energy resolution was about 1.0 eV as measured in full-width at half-maximum of the zero-loss peak. The Fe $L_{2,3}$ edges spectra were taken from the crystals of $\sim 25$~nm thick under diffraction mode with a collection angle of $\sim 3.9$~mrad.

First, we use CBED to accurately measure the low-order electron structure factors of \bfa, as shown in Figs.~\ref{fig:CBED}(a)-\ref{fig:CBED}(c) and Table S1 in Supplemental Material \cite{SI}
for $x=0$. The higher-order structure factors (which are dominantly contributed by atomic position and core charge) can be obtained from x-ray diffraction \cite{Zuo,Wu99,Wu,Zhu} or density-functional theory (DFT) electronic structure calculation without losing significant accuracy \cite{Wu,Zhu}. Then, we run the multipole refinement over the combined CBED and x-ray or DFT data (the DFT data are used here) to yield three-dimensional charge density distribution. The Co substitution disorder is treated in the multipole refinement using a \emph{posterior} virtual crystal approximation \cite{SI}. Since the total density map is dominated by the core electrons, the valence electron distribution is presented by the difference map between the aspherical crystal charge density and superimposed spherical atomic charge density \cite{Zheng}.

%figure1
\begin{figure}[t]
%\vspace{1cm}
\includegraphics[width=0.95\columnwidth,clip=true,angle=0]{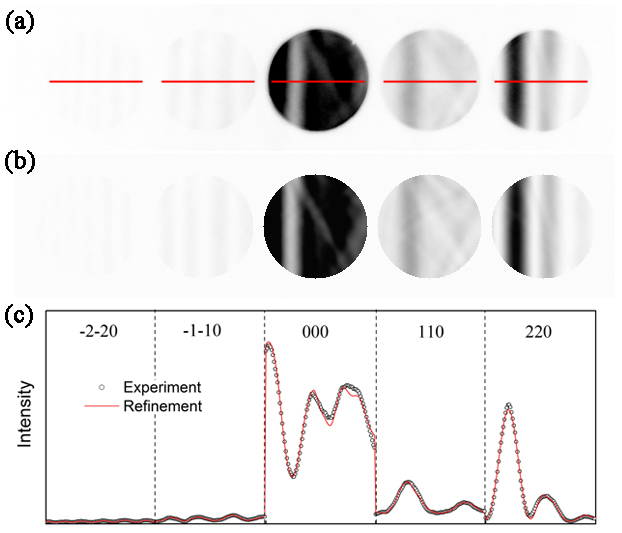}
\caption{\label{fig:CBED}\textbf{Measurement of the low-order electron structure factors using CBED.} (a) Experimental energy-filtered CBED pattern of BaFe$_2$As$_2$ showing the $110$ systematical row (reflection of $\bar{2}\bar{2}0$, $\bar{1}\bar{1}0$, $000$, $110$, and $220$) at room temperature. (b) Calculated pattern using the dynamical Bloch wave method. (c) Line scans of the intensity profile from the experimental pattern (open circle) and calculated one (red line) after the structure factor refinement.}
\end{figure}

\begin{figure}[t]
%\vspace{1cm}
\includegraphics[width=0.95\columnwidth,clip=true,angle=0]{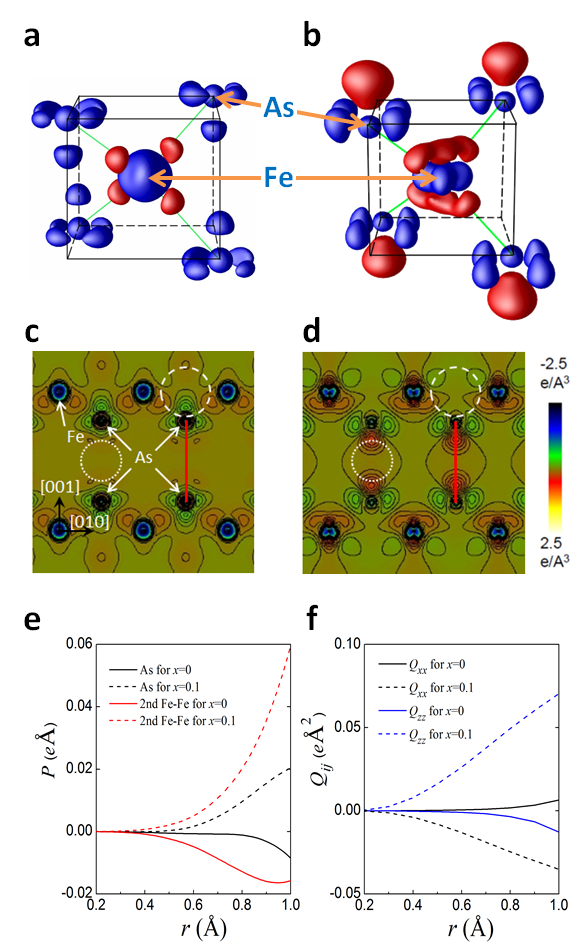}
\caption{\label{fig:3d}\textbf{Valence electron density map in \bfa\ } for (a) $x=0$ and (b) $x=0.1$ as well as the two-dimensional map in the (100) plane for (c) $x=0$ and (d) $x=0.1$. The isovalues of difference charge density on the isosurface are 0.1 $e$/{\AA}$^3$ (red) and $-0.1$ $e$/{\AA}$^3$ (blue), respectively. The color legend indicates the magnitude of the charge density and the contour plot has an interval of 0.05 $e$/{\AA}$^3$. The red line denotes the shortest As-As distance between the FeAs layers. (e) The radius dependence of the out-of-plane dipole moment inside the sphere centered at the As anion (black solid and dashed lines) and inside the sphere centered at the middle point of the next-nearest Fe-Fe bond illustrated as the white dashed circle in (c) and (d) (red solid and dashed lines).  (f) The radius dependence of the quadrupole moments inside the sphere centered at the Fe anion.}
\end{figure}

The resulting room-temperature three-dimensional difference charge density in \bfa\ is shown in Fig.~\ref{fig:3d}(a) for $x=0$ and  Fig.~\ref{fig:3d}(b) for $x=0.1$ the near optimally doped sample. The essential crystal structure of \bfa\ is the FeAs trilayer: The As anions form a two-dimensional network of edge-shared tetrahedra; the center of each tetrahedron is occupied by an Fe cation, forming the conducting Fe square lattice \cite{Sefat}. The valence electron distribution is characterized by a significant increase in both the charge quadrupole around the Fe cations (more aspheric with inversion symmetry) and the charge dipole around the As anions (more aspheric without inversion symmetry) up Co substitution, echoing the significant increase in $T_c$, as elaborated below.

The charge quadrupole moments inside a sphere centered at the Fe cation are given by $Q_{ij}=e\int \Delta \rho (\vec{r})[3(\vec{r}-\vec{R}_{{\rm Fe}} )(\vec{r}-\vec{R}_{{\rm Fe}} )-I(\vec{r}-\vec{R}_{{\rm Fe}} )\cdot (\vec{r}-\vec{R}_{{\rm Fe}} )]_{ij} d\vec{r} $, where $\vec{r}$ and $\vec{R}_{{\rm Fe}}$ are the positions of the electrons and the Fe ion, respectively, \textit{e} the electron charge, $\Delta \rho (\vec{r})$ the difference electron density at $\vec{r}$, and \textit{I} the 3x3 identity matrix. Its diagonal \cite{note:diagonal} quadrupole moments $Q_{xx} $ and $Q_{zz} $ as a function of the radius of the Fe sphere, $r$, are shown in Fig.~\ref{fig:3d}(f). Upon Co substitution, $Q_{xx} $ decreases and $Q_{zz} $ increases, both by a remarkable amount, indicating a strong interorbital charge transfer of about 0.25 electrons from the in-plane Fe $3d$ orbitals ($x^2-y^2$ and $xy$) to the out-of-plane Fe $3d$ orbitals ($z^2$, $xz$, and $yz$). As a result, the electronic structure of \bfa\ would be more three-dimensional upon Co substitution, in agreement with angle-resolved photoemission spectroscopy (ARPES) \cite{Thirupathaiah}. This observation points to an unusual character of iron-based superconductors, namely the rather weak crystal-field splitting of the Fe $3d$ orbital levels. Usually, the layered structure hosts a strong crystal-field splitting between the in-plane and out-of-plane $d$ orbitals; therefore, charge fluctuation between them is suppressed.
However, in iron-based superconductors the largest $3d$ splitting is about 0.5 eV, which is comparable to the interorbital electron hopping strength \cite{Yin:PRL99}. Therefore, the Fe-orbital degree of freedom is nearly unquenched and strong orbital fluctuation involving all the five Fe $3d$ orbitals can take place.

To facilitate the discussion about the anion polarization, we present the difference charge density maps in the (100) plane in Figs.~\ref{fig:3d}(c) and \ref{fig:3d}(d) for $x=0$ and $0.1$, respectively. Apparently, when \textit{x} is increased from 0 (\textit{T${}_{c}$} = 0 K) to 0.1 (\textit{T${}_{c}$} = 22.5 K), remarkably more electrons (more reddish area in Fig.~\ref{fig:3d}d) appear in between the layers of arsenic anions. The number of valence electrons within the interstitial sphere (illustrated as the white dotted circles in Figs.~\ref{fig:3d}c and ~\ref{fig:3d}d) dramatically increases from 0.008 at $x=0$ to 0.114 at $x=0.1$ when the diameter is half of the interlayer As-As distance, while the As-As overlap remains weak. This means that Co substitution induces a substantial shift of the electron cloud of the arsenic anion. This shift, i.e., the anion polarization, can be quantified by the charge dipole inside the sphere centered at the arsenic anion, $\vec{P}=e\int \Delta \rho (\vec{r})(\vec{r}-\vec{R}_{{\rm As}} )d\vec{r} =(0,0,P)$, where $\vec{R}_{{\rm As}} $ is the position of the arsenic anion. Its out-of-plane component as a function of the radius of the sphere is shown in Fig.~\ref{fig:3d}(e) (black solid and dashed lines). It increases substantially from $-0.009$ \textit{e}{\AA} for $x=0$ to 0.020 \textit{e}{\AA} for $x=0.1$ when the radius is 1.0~{\AA} (half of the Fe-As distance is 1.2~{\AA}).

The considerable accumulation of valence electrons in between the FeAs layers at optimal doping (Fig.~\ref{fig:3d}d) may lead to the impression that the interlayer As-As bonding (red line in Fig.~\ref{fig:3d}d) appears. This is reminiscent of the pressure experiment on CaFe$_2$As$_2$, where the interlayer As-As distance collapses from 3.78~{\AA} to 3.0~{\AA}, accompanied by the vanishing of the Fe spin moment and the emerging of superconductivity \cite{Kreyssig}. By contrast, the interlayer As-As distance in \bfa\ for $x = 0 - 0.1$ remains nearly unchanged at 3.78 {\AA} (ref. \onlinecite{Sefat}), which is too distant for a real bonding to occur. Indeed, we found that the number of electrons is still vanishing at the middle of the As-As ``bond'' (the red line in Figs.~\ref{fig:3d}c and~\ref{fig:3d}d), indicative of little overlap of the As-As atoms along the $c$ axis. We thus conclude that the shift of the anions' electron cloud in \bfa\ is caused by the Fe electrical field, not by the As-As bonding. \red{In addition, since the structural change induced with Co doping is rather small, it is ruled out as a driving force for the observed large change in the electron redistribution \cite{SI}.}

The strong anion polarization should in turn polarize the Fe-Fe bonds. We notice that the middle points of the nearest-neighbor (NN) and next-nearest-neighbor (NNN) Fe-Fe bonds are distinct in symmetry: the former is an inversion symmetry center but the latter is not. Therefore, the NNN electron hopping should be much more affected by anion polarization than the NN one. We calculated the electric dipole moment $\vec{P}$ within the sphere centered at the middle point of the NNN Fe-Fe bond (the white dashed circles in Figs.~\ref{fig:3d}c and ~\ref{fig:3d}d); this is the in-Fe-plane point directly under an As anion. We found that its out-of-plane component changes substantially in a similar manner as the anion polarization, e.g., from $-0.016$ for $x=0$ to $0.060$ \textit{e}{\AA} for $x = 0.1$ when the diameter is half of the NNN Fe-Fe distance, being consistent with dipole-dipole interaction.
This effect may be relevant to iron-based superconductivity, since the NNN Fe-Fe superexchange, $J_2$, was correlated with $T_c$ \cite{MaFJ}.

%\textsl{Discussion.---}%
The remarkable simultaneous redistribution of the valence electrons around the As and Fe atoms should be reflected in a large change in phonon modes that involve both Fe and As atoms. Indeed, the infrared-active in-plane \textit{E${}_{u}$} mode observed at about 258 cm${}^{-1}$ in BaFe${}_{2}$As${}_{2}$ (above the magnetic transition) disappears in optimally Co-doped samples \cite{Akrap,Tu,Schafgans}. To better understand the nature of the lattice vibrations, \textit{ab} \textit{initio} calculations were performed using the direct method for the zone-center phonons
\cite{SI}.
We found that the atomic character of this particular in-plane mode is almost evenly distributed between the Fe and As atoms (55\% Fe, 45\% As). However, this should simply lead to line broadening, not the almost total extinction of this feature.~For this reason, the change in the~atomic charge distribution and the commensurate increase in the screening from electron polarization may be an explanation.

To understand these surprising observations, we note that the As${}^{3-}$ polarization, \textit{P}, is mainly caused by the asymmetrical positioning of Fe cations around the As${}^{3-}$ anion along the \textit{z} axis. Thus, \textit{P} can be generally expressed in the polynomials of Fe/Co charge $n_{i}$: $P=\beta _{1} \left\langle n_{i} \right\rangle +\beta _{2} \left\langle n_{i} ^{2} \right\rangle +\cdots $. Since $\left\langle n_{i} \right\rangle $, the average Fe/Co charge monopole, remains unchanged upon Co substitution, providing Ba and As remain $2+$ and $3-$, respectively, a significant change in $\left\langle n_{i} ^{2} \right\rangle $ must be responsible for the observed substantial change in \textit{P}. This implies a significant change in the degree of electronic correlation, which we found is unsurprisingly beyond the capability of local density density approximation of density functional theory \cite{SI}. Furthermore, the nonlinearity of anion polarization is enhanced by intersite charge fluctuation in the Fe plane. For example, the charge fluctuation between two Fe sites \textit{i} and \textit{j}, say one electron hops from \textit{j} to \textit{i}, satisfies the inequality $\left(n_{i} +1\right)^{2} +(n_{j} -1)^{2} >n_{i} ^{2} +n_{j} ^{2} $ and changes \textit{P} by 2\textit{$\beta$}${}_{2}$. The nonlinearity of the anion polarization was found in a recent one-orbital theoretical study \cite{Sawatzky,Berciu}. We expect that a multiorbital model in which the Fe orbital fluctuation can ``resonate'' with the anion polarization is necessary to explain the observed extraordinary charge response to the Co doping.

%\textsl{EELS.---}%
\begin{figure}[t]
%\vspace{1cm}
\includegraphics[width=0.9\columnwidth,clip=true,angle=0]{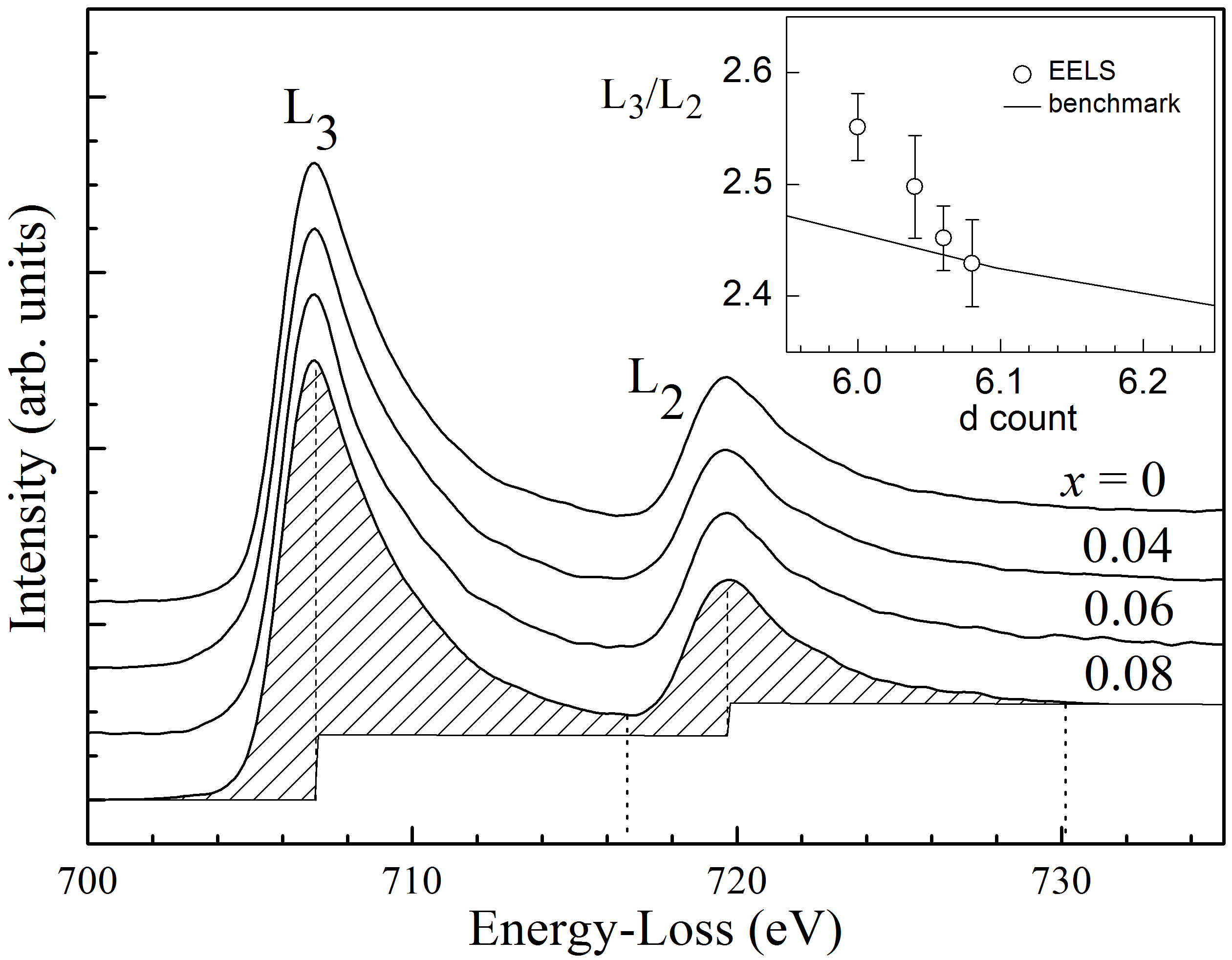}
\caption{\label{fig:EELS}\textbf{Co concentration dependence of Fe $L_{2,3}$ edges EELS} for \bfa\ for $x = $ 0, 0.04, 0.06, and 0.08. The inset: The measured L$_3$/L$_2$ white line ratio (circles) changes with $6+x$, the count of $3d$ electrons assuming Co$^{3+}$, considerably more than \red{the benchmark} (line, see the text), indicating that Co substitution introduces charge carriers.}
\end{figure}

Since $\left\langle n_{i} ^{2} \right\rangle $ and intersite charge fluctuation are usually enhanced by charge doping away from the stoichiometric limit $x=0$, the above analysis implies that substituting Co for Fe effectively dopes charge carriers into the Fe planes. \red{This Co-doping effect is actually a controversial issue   \cite{SI,Thirupathaiah,Rullier-Albenque,Konbu,Berlijn,Bittar,Merz,Wadati}}.
To get more insight, we study the room-temperature Co concentration dependence of Fe $L_{2,3}$ edges EELS. As shown in Fig.~\ref{fig:EELS}, two peak areas marked as Fe $L_2$ and $L_3$ appear at about 719.6 and 707 eV, respectively. They originate from $2p$-$3d$ dipole transitions with the well-separated spin-orbit-splitting $2p$ states $2p_{1/2}$ ($L_2$) and $2p_{3/2}$ ($L_3$). The spectra, not exhibiting any multiplet peak structures, are similar to those for metallic Fe, but much different from those for the iron oxides known as Mott insulators. Upon Co substitution, the $L_3/L_2$ ratio drops from 2.55 at $x = 0$ to 2.43 at $x = 0.08$. \red{Note that this drop implies a slight decrease of the Fe local moment with Co substitution \cite{Thole,Pease}, in agreement with x-ray emission spectroscopy \cite{Gretarsson}.} To properly evaluate how significant this change is, \red{we employ the theoretical $L_3/L_2$ ratio for $3d$ transition-metal compounds \cite{Thole,Pease} to \emph{benchmark} it.} As shown in the inset of Fig.~\ref{fig:EELS}, the measured $L_3/L_2$ ratio as a function of $n=6+x$ (the nominal count of Fe $3d$ electrons assuming that one Co atom dopes one electron) \red{is comparable to but changes noticeably more than the benchmark}, meaning that the observed change is substantial. Our EELS data thus favor the picture that Co substitution introduces charge carriers.

Our observations suggest a unified picture for the orbital fluctuation  \cite{Kontani,Yanagi} and electronic polaron \cite{Sawatzky,Berciu} physics in \bfa: The Co substitution in \bfa\ effectively dopes charge carriers into the system. This enhances intersite charge fluctuation in the Fe plane and $\left\langle n_{i} ^{2} \right\rangle$. In accordance, the nonlinear electronic polarization of the As anions is strengthened. In turn, the enhanced anion polarization reduces the electron-electron repulsion in the Fe plane, which enlarges $\left\langle n_{i} ^{2} \right\rangle$ and induces charge redistribution among the Fe $3d$ orbitals via the As dipole-Fe quadrupole interaction. This positive feedback gives rise to a purely electronic mechanism for boosting Fe-orbital fluctuation.
%\red{The present results also suggest that the fluctuating Fe quadrupole moment $Q_{xx-yy}$ considered important to superconductivity \cite{Fernandes,Goto,Yoshizawa,Kontani,Yanagi} could be induced by \emph{in-plane} optical oscillation of the anions' electron cloud \cite{note:diagonal}.}

In summary, we have studied the Co concentration dependence of the valence electron distribution in \bfa\ using advanced electron probes. Our data provide the first experimental evidence that the \red{charge redistribution among the iron $3d$ orbitals} is strongly coupled with anion polarization.
The mechanism that leads to superconductivity in iron pnictides and chalcogenides will only be fully understood once the charge, spin, orbital, lattice, and anion polarization are all together considered in a consistent theory.

We are grateful to the late Myron Strongin for stimulating discussions throughout this project. We thank Laurence D. Marks for helpful communication. Work at Brookhaven National Laboratory was supported by the U.S. Department of Energy, Office of Basic Energy Science, Division of Materials Science and Engineering, under Contract No. DE-AC02-98CH10886. 
Work at Institute of Physics, CAS was supported by the National Science Foundation of China (Grant Nos. 11190022, 11004229) and the National Basic Research Program of China 973 Program (Grant No. 2011CBA00101).

\textsl{Note added in proof.}---The relevance of the orbital fluctuation between the Fe in-plane and out-of-plane orbitals to higher $T_c$ was recently shown in a theoretical paper by Onari, Yamakawa, and Kontani \cite{Onari}.


\begin{thebibliography}{99}
\bibitem{note:diagonal} \red{Constrained by the symmetry of space group $I4/mmm$ used in our multipole refinement, the Fe dipole moment and off-diagonal $Q_{ij}$ are vanishing and $Q_{xx}=Q_{yy} $, while the anion in-plane dipole moments are zero.}
\end{thebibliography}
\end{document}